\documentclass[pra,aps,tightenlines,showpacs,twocolumn]{revtex4}
\usepackage{epsfig}
\def\bra#1{\langle #1 |}
\def\ket#1{| #1\rangle}
\def\Tr{{\rm{Tr}}}

\begin{document}

\title{Inferring superposition and entanglement in evolving systems
from measurements in a single basis}

\author{Bella Schelpe, Adrian Kent$^*$, William Munro and Tim Spiller}
\affiliation{Hewlett-Packard Laboratories, Filton Road,
Stoke Gifford, Bristol BS34 8QZ, U.K.}

\date{March 2003 (revised)}

\begin{abstract}
We discuss what can be inferred from measurements on evolving one and
two qubit systems using a single measurement basis at various times.
We show that, given reasonable physical assumptions, carrying out
such measurements at quarter-period intervals is enough to demonstrate
coherent oscillations of one or two qubits between the relevant
measurement basis states. One can thus infer from such measurements
alone that an approximately equal superposition of two measurement
basis states has been created during a coherent oscillation experiment.
Similarly, one can infer that a near maximally entangled state of two
qubits has been created part way through an experiment 
involving a putative SWAP gate.  These results apply even if the 
relevant quantum systems are only approximate qubits.  We discuss 
applications to fundamental quantum physics experiments and 
quantum information processing investigations.
\end{abstract}

\maketitle
\section{Introduction}

It is well known from tomography that given many copies of a quantum state,
measuring enough observables allows reconstruction of the state \cite{tomography}.
However, the relevant quantum tomographic schemes generally require 
measurements in more than one basis. In many systems this doesn't 
present any difficulties, for example polarisation measurements on 
photons are easily carried out in any basis. However for many
condensed matter qubit systems, there is usually just one natural measurement
basis, often defined to be the computational basis or, in spin-1/2
notation, the $z$-basis: $|0\rangle = |\uparrow_{z}\rangle$ and
$|1\rangle = |\downarrow_{z}\rangle$.  This is the motivation for this
paper. We would like to know what, if anything, can be inferred about
the evolution of a one or two qubit system if measurements on each
qubit are restricted to a single basis, but can be made at different
times. Specifically, we are interested in the problems of demonstrating
coherent oscillations between $\ket{0}$ and $\ket{1}$ in an
experimental arrangement that is meant to function as a quantum
NOT gate, and between $\ket{0}\ket{1}$ and $\ket{1}\ket{0}$ in an
arrangement that is meant to function as a quantum SWAP gate.
Our goal is thus less ambitious than full quantum state tomography.
However, the experimental simplicity of our approach could be
used to infer evidence for the fundamental quantum phenomena of 
superposition and entanglement in new candidate qubit systems, 
which are currently not amenable to full tomography \cite{colls}. 
Other approaches such as ''entanglement witnesses'' have been proposed 
to test whether a system  is entangled or now\cite{lewenstein,guhne}. 
These generally however require measurements outside the computational 
basis that we wish to restrict our attention too.

Consider first the NOT gate.  The standard coherent oscillation
scenario is ideally realised by arranging for the spin qubit
Hamiltonian to be of the form $H = - \frac{\Gamma}{2} \sigma_{x}$,
with $\hbar$ set to $1$. Inputting either computational basis state
($\sigma_z$ eigenstate), the probabilities of finding the system in
these basis states oscillate coherently with period $2 \pi
\Gamma^{-1}$. With a dimensionless time $T \equiv \Gamma t$, a single
oscillation is complete at $T= 2 \pi$. A NOT gate is effected by the
action of $H$ for a time $T = \pi$ and a root NOT by acting for $T =
\pi/2$.

The usual approach to inferring that a superposition 
$\ket{0} + \ket{1}$ really was created at $T = \pi/2$ in an 
experiment is to plot out the sinusoidal oscillations of the 
computational basis measurement probabilities in detail.  
This is done by preparing, evolving for a small fraction of 
$T = 2 \pi$, measuring in the computational basis, repeating 
many times to build up statistics of $p(0)$ and $p(1)$ for 
this value of $T$, then incrementing $T$ and repeating the 
whole procedure until one reaches $T= 2 \pi$.

Coherent oscillation of the quantum state
appears much the simplest and likeliest explanation of a
sinusoidal oscillation in the measurement probabilities.
One might worry, though, that some other evolutions could
also be consistent with the data.  For example, measurement probabilities
$p(0) = p(1) = 1/2$ at $\pi/2$ would also be obtained if the system
generated a mixed state $ \ket{0} \bra{0} + \ket{1} \bra{1}$,
rather than the desired pure state $\ket{0} + \ket{1}$,
at this point.  Conversely, one might wonder whether carrying out
measurements at many evolution times is necessary or even helpful
for inferring the state at $\frac{\pi}{2}$.

A further issue in inferring superposition is that real world
plots will at best only be approximately consistent with ideal coherent
oscillation: for example, the experimental oscillation amplitudes will
decay over time.  Moreover, the relevant quantum system may only
be an approximate qubit. Textbook discussions of quantum 
computing often assume the qubits are
exact, in that they correspond to two-dimensional quantum systems such
as the polarisation states of a single photon.  However, 
real world qubits may be approximate, corresponding to
systems which live in a larger Hilbert space but have parameters set
so that two of the states have very small transition probabilities to
any of the rest. One would ideally like to be able to estimate how close 
the state generated under these conditions is to the superposition
obtained by an ideal evolution.

In this paper we show that, in fact, the existence of coherent
superpositions can be inferred simply by carrying out computational
basis measurements every quarter-period, without a full sinusoidal
oscillation plot.  This is true for both the NOT and SWAP gate
evolutions, and even for approximate qubits and imperfect evolutions.
In all cases a reasonably simple bound on the fidelity to an
ideal superposition state can be obtained.  Our arguments require
physically reasonable and fairly minimal assumptions, which are
necessary to infer coherent superposition even if measurements are
made at many different evolution times.  Our procedure thus simplifies
the experimenter's task considerably, while providing evidence 
for the results inferred.

The experimental motivation for considering approximate as well as exact 
qubits comes from condensed matter systems where the 
qubits are often not exact. Two interesting condensed matter
approximate qubits are based on superconductivity.  The computational
basis states can either be two charge states of a microscopic
superconducting island (or box) differing by a single Cooper pair
\cite{shni}, or two flux states of a closed
superconducting ring \cite{bocko,mooji}.

Such systems are particularly interesting because they form one of a
number of promising routes towards realising condensed matter qubits for
quantum computing. Coherent oscillations have already
been seen with an approximate charge qubit \cite{nakamura}, similar results 
have been seen with related charge-phase systems \cite{vion,yu,martinis} and
very recently two charge qubits have been coupled \cite{pashkin}, 
leading to the possibility of entanglement. Flux systems may also be 
useful qubits and, in addition, these are very interesting from the fundamental
quantum physics viewpoint. Two flux states of a superconducting ring,
differing by the flux quantum $\Phi_{0} = h/2e$, correspond to
macroscopically distinct circulating currents in the ring, so a
superposition of two such states would be a real analogue of
Schr\"{o}dinger's cat.  Flux experiments \cite{friedman,wal} have provided
evidence for flux superpositions in the frequency domain, in effect
through spectroscopy, although these experiments did not exhibit time
domain oscillations. Various experiments to demonstrate coherent flux
oscillations are underway, and the first results are emerging \cite{chior}.
Flux or current oscillations will be recognised as clear evidence for 
Schr\"{o}dinger cat superpositions (in the same way that charge 
oscillations \cite{nakamura} have been recognised as showing clear evidence 
for charge superpositions). The single approximate qubit results we present 
here are particularly relevant for the ongoing flux oscillation experiments.

For two qubit systems there are a number of potential gates that can be
investigated. Here however we restrict our attention to the SWAP and
root SWAP gates. These are useful gates to consider, because as is well 
known, universal quantum computations can be carried out 
by combining root SWAP with single qubit gates.
A swapping evolution (involving just two of the four basis states of a
two qubit system) is a natural one for various superconducting systems
\cite{pashkin,strauch,example}. For a system of two spins, $a$ and $b$, 
a swapping evolution can be realised with an interaction
Hamiltonian of the Heisenberg form
$H = \frac{J}{4}\underline{\sigma}_{a}.\underline{\sigma}_{b}$.
This interaction is relevant for electron spins in coupled
quantum dots \cite{div1} and other spin realisations of qubits in condensed
matter systems. The computational basis states $\ket{0}\ket{1}$ and
$\ket{1}\ket{0}$ swap after a dimensionless time $Jt\equiv T = \pi$.
After a time $T = \pi/2$ a root SWAP gate is effected \cite{div1}.
Approximate implementations of root SWAP have also been proposed for
other condensed matter qubit systems, electron on helium
\cite{lea} and Coulomb coupled quantum dots \cite{jefferson}
where the approximate qubits are fictitious spins, so the swapping 
evolution is relevant for numerous qubit realisations.

\section{Physical assumptions}

Our fundamental assumption is that, in any experiment we consider,
we can characterize the Hilbert space corresponding to the
quantum system(s) of interest, and distinguish
system degrees of freedom from those corresponding to the rest of
the experimental apparatus, thermal radiation, and the rest
of the outside world, all of which we collectively refer to
as the environment.  So the total Hilbert space ${\cal H}_{\rm total}$
can be factored as
$$
{\cal H}_{\rm total} = {\cal H}_{S} \otimes {\cal H}_{E} {\rm~or~}
{\cal H}_{S_1} \otimes {\cal H}_{S_2} \otimes {\cal H}_{E} \, ,
$$
for a one or two qubit system respectively.
Exact qubits correspond to two dimensional spaces ${\cal H}_{S}$
and ${\cal H}_{S_i}$ with an orthonormal basis $\ket{0}, \ket{1}$.
For approximate qubits we can without loss of generality consider 
countably infinite dimensional spaces with a basis 
$\ket{0}, \ket{1}, \ket{2} , \ldots$, with the
two distinguished states $\ket{0}$ and $\ket{1}$ defining the qubit
computational basis.

We assume that the system-environment interactions are such that
we can initialize the system in any computational basis state without
materially affecting the environment and can treat the
environment as effectively constant during subsequent
evolutions.  More precisely:

(i) the state of the experiment at any time can be
characterized by a density matrix $\rho_S$ or $\rho_{S_1 \otimes S_2}$
describing (only) the system state.

(ii) the evolution of the system between any times $t$ and $t' > t$ is
described by a quantum operation ${\cal E}_{t, t' } $ which
is trace-preserving, convex linear on density matrices and
completely positive.

(iii) the operations ${\cal E}_{t, t' }$ is independent of the initial
state of the system and is the same in each experimental run.

We turn now to the detailed analyzes of the various qubit evolutions and
what may be inferred about superposition and entanglement from measurements.

\section{An exact qubit}

There is a particularly simple strategy for inferring superpositions
in the case of an experiment testing a prototype NOT gate by
attempting to produce coherent oscillations of a single exact qubit.
We run the experiment setting the initial system state to be
either of the two computational basis states $\rho_0 = \ket{0}
\bra{0}$ and $\rho_1 = \ket{1} \bra{1}$.  Assume that we have
identified, either theoretically or empirically, a time $t$ such that
${\cal E}_t \equiv { \cal E}_{0,t}$
implements an approximate NOT operation.  We can quantify the degree
of approximation by carrying out a series of experiments in
which computational basis measurements are performed at time $t$ and
estimating the probabilities of obtaining $0$ and $1$ from the
states evolved from $\rho_1$ and $\rho_0$. We write
\begin{eqnarray}
\label{notprobs}
p_{i}^{j} &=& \bra{i} { \cal E}_{t} ( \rho_{j} ) \ket{i} \,.
\end{eqnarray}

To proceed we need the following definitions and results\cite{ncninetwo}.
The {\it trace distance} between states $\rho$ and $\rho'$
is defined as
$ D ( \rho , \rho' ) = \frac{1}{2} \Tr | \rho - \rho' | \, , $
where $ | A | = ( A^{\dagger} A )^{\frac{1}{2}} $ is the positive square
root of $A^{\dagger} A $.   It has the following properties.
First, $D$ is a metric: $ D( \rho , \rho' ) = 0$
if and only if $\rho = \rho'$, $ D ( \rho , \rho' ) = D ( \rho' , \rho )$
and $ D ( \rho , \rho'' ) \leq D ( \rho , \rho' ) + D ( \rho' , \rho'' ) $.
Second, $D$ is non-increasing under
trace-preserving quantum operations:  $ D ( {\cal E} ( \rho ) ,
{\cal E} ( \rho' ) ) \leq D ( \rho , \rho' )$.
Third, the {\it fidelity}
$F ( \rho , \rho' ) = \Tr ( \rho^{\frac{1}{2}} \rho'
\rho^{\frac{1}{2}}  )$
and $D$ obey
\begin{eqnarray}
\label{fiddist}
 1 - F^2 ( \rho , \rho' ) \leq D ( \rho, \rho' ) \leq
( 1 - F(\rho, \rho') )^{\frac{1}{2}} \, .
\end{eqnarray}
When $\rho$ is a pure state we have a stronger lower bound:
\begin{eqnarray}
\label{fiddistpure} 1-F(\rho,\rho')\leq D(\rho,\rho').
\end{eqnarray}
We will need a further result which is that when $\rho$ is a given pure
state, $F(\rho,\rho')$ is the same for all $\rho'$ lying on a disc
orthogonal to the radius joining the centre of the Bloch sphere to
$\rho$.

This can be seen by considering $\rho = \ket{0}\bra{0}$. The pure 
states on a given disc are
$\cos\frac{\theta}{2}\ket{0}+e^{i\phi}sin\frac{\theta}{2}\ket{1}$.
The fidelities of all these states to $\ket{0}\bra{0}$ are
$\cos^2\frac{\theta}{2}$. Since all mixed states on the disc are
mixtures of these pure states and fidelity between a mixed and
pure state is linear in the mixed state, all states on the disc
have fidelity $\cos\frac{\theta}{2}$ to $\ket{0}$.

 If our system really is an exact qubit, we find that $ p_0^1
+ p_1^1  = 1 = p_0^0 + p_1^0$. Geometrically, ${ \cal E}_{t} (
\rho_1 )$ lies in the intersection of the Bloch sphere with a
plane at height $2 p_0^1 - 1$ above the equator and parallel to
it, while ${ \cal E}_{t} ( \rho_0 )$ lies in the intersection of
the sphere with a parallel plane $2 p_1^0 - 1$ below the equator.
Since the trace distance is half the Euclidean distance within the
Bloch sphere, we have that 
\begin{eqnarray} \label{dist}
 D ( { \cal E}_{t} ( \rho_1 ) , { \cal E}_{t} ( \rho_0 ) ) 
 \geq (  p_0^1 +p_1^0 - 1 ) \, .
\end{eqnarray}

As the trace distance is non-increasing during quantum evolution,
the two qubits evolved from $\rho_0$ and $\rho_1$ must always have
been separated by at least this distance before time $t$.  We can
picture the qubit evolution as defining that of a perhaps
contracting rod (whose endpoints are the qubits) moving inside the
Bloch sphere. At some time before $t$ during the evolution, this
'rod' must be parallel to the equatorial plane. 
The condition (\ref{dist}) defines a cylinder of Euclidean 
radius $p^1_0 +p^0_1 -1$ outside which at least one of the 
qubits must lie at this time.  The state most distant from 
any pure state on the equator must therefore enter a 
volume bounded by the Bloch sphere and a disc normal to the radius 
to some equatorial point and subtended by a geometric angle defined
by $\cos \theta = p_0^1 + p_1^0 - 1$.
Using equation (\ref{fiddistpure}) and the above result 
that all $\rho'$ lying on a disc orthogonal to the relevant radius 
will have the same fidelity, we can see that 
the fidelity of this qubit to a maximally superposed state
of the form $\ket{0} + e^{i \phi } \ket{1}$ must be no less than $
( \frac{ p_0^1 + p_1^0 }{2} )$. Hence we can infer
the creation of a near-maximally superposed state during the
experiment --- though note that this argument does not identify
which superposed state was approximated, or when.

\section{An approximate qubit}

Now suppose our system is only approximately characterized by
a qubit.  Clearly, any argument for coherent oscillations will
need some measure of how closely the evolved state resembles
a qubit.  For any given state $\rho$ that can be repeatedly
prepared, we can obtain a good measure as follows.
Define $P$ to be the projection onto the two-dimensional qubit
space, and note that if $p_0^{\rho}$ and $p_1^{\rho}$ are the
probabilities of getting outcomes $0$ and $1$ when measuring
in the computational basis, we can obtain
$\Tr ( P \rho ) = p_0^{\rho} + p_1^{\rho}$.
Let $ \rho_P = P \rho P /  \Tr ( \rho P ) $ be the normalized
density matrix given by projecting $\rho$ into the qubit space.
Then
$F ( \rho_P , \rho ) = ( \Tr ( P \rho ) ) =
( p_0^{\rho} + p_1^{\rho} )$
and so we have
$$
D ( \rho_P , \rho ) \leq ( 1 - p_0^{\rho} - p_1^{\rho} )^{\frac{1}{2}} \, .
$$

We can now proceed as previously and obtain
estimates for the probabilities (\ref{notprobs}) associated
with an approximate NOT evolution, giving us that
\begin{widetext}
\begin{eqnarray}
D ( { \cal E}_{t} ( \rho_1 ) , { \cal E}_{t} ( \rho_0 ) ) & \geq &
D ( { \cal E}_{t} ( \rho_1 )_P  , { \cal E}_{t} ( \rho_0 )_P ) -
D ( { \cal E}_{t} ( \rho_0 )  , { \cal E}_{t} ( \rho_0 )_P ) -
D ( { \cal E}_{t} ( \rho_1 )  , { \cal E}_{t} ( \rho_1 )_P ) \\
\nonumber
\vspace{-0.5cm}
& \geq &
 ( q_0^1 + q_1^0 - 1 ) - ( 1 - p_0^0 - p_1^0 )^{\frac{1}{2}}
- ( 1 - p_0^1 - p_1^1 )^{\frac{1}{2}} \, ,
\end{eqnarray}
\end{widetext}
where $q_0^1 = p_0^1 / ( p_0^1 + p_1^1 )$ and
$q_1^0 = p_1^0 / ( p_0^0 + p_1^0 )$.

To complete the argument, we need some way of bounding the possible
deviation from an exact qubit at a point when we expect a
near-maximally superposed state to have been created.  A general bound
of the form $ \Tr ( P {\cal E}_{t'} (\rho_i ) ) \geq 1 - \delta$ for
all times $t' < t$ and for $i=0,1$
would suffice.  However, getting good
empirical evidence for such a bound would require carrying out a
series of measurements at many times between $0$ and $t$, which
would require the same amount of experimental labour as plotting
sinusoidal oscillations, and would still leave the worry that the
deviation might have been greater at some unmeasured intervening
point.

A more watertight procedure is to identify empirically
a time $t_{1/2} < t $ (for coherent oscillations
we expect $t_{1/2} \approx t/2$) such that, if we write
$
r_i^j = \bra{i} { \cal E}_{t_{1/2}} ( \rho_j ) \ket{i} \, ,
$
we have
$
r_0^0 \, ,  r_1^0 \, ,  r_0^1 \, , r_1^1  \approx 1/2 \, .
$
We can then directly argue that the evolution came close to
a maximally superposed state at $t_{1/2}$ as follows.

Writing $\sigma_i = { \cal E}_{t_{1/2}} ( \rho_i )$
and $\sigma^P_i = P  \sigma_i  P / \Tr ( \sigma_i P ) $
for $i = 0,1$ we have
$
F ( \sigma^P_i , \sigma_i ) = ( \Tr ( P \sigma_i ) )
= ( r_0^i + r_1^i ) \,
$
and
\begin{widetext}
\begin{eqnarray}
D ( \sigma^P_0 , \sigma^P_1 ) & \geq & D ( \sigma_0 , \sigma_1 ) -
D ( \sigma_0 , \sigma^P_0 ) - D ( \sigma_1 , \sigma^P_1 )  \\ \nonumber
&\geq &  ( q_0^1 + q_1^0 - 1 ) -  ( 1 - p_0^0 - p_1^0 )^{\frac{1}{2}}
- ( 1 - p_0^1 - p_1^1 )^{\frac{1}{2}}
- ( 1 - r_0^0 - r_1^0 )^{\frac{1}{2}}
- ( 1 - r_0^1 - r_1^1 )^{\frac{1}{2}} \, .
\end{eqnarray}
\end{widetext}
We will write the right hand side of this last inequality as $1 -
\epsilon'$ and let $s_j^i = r_j^i / ( r_0^i + r_1^i )$, with $\Delta
s_j^i = s_j^i - 1/2$.  A little Bloch sphere trigonometry shows that
at least one of the states $\sigma^P_i$ must lie in a cap, centred on
a maximally superposed state, whose Euclidean height is no more than
$$
\delta =
 1 - \sqrt{( 1 -  \epsilon' )^2 -  (\Delta s_0^1 + \Delta s_1^1 )^2 }  \, .
$$
The relevant $\sigma^P_i$ is separated from the maximally superposed state
$ \ket{ \psi_M }$ by trace distance no more than
$
( \frac{\delta^2}{4} + ( \Delta s_0^i )^2 )^{\frac{1}{2}} \, ,
$
and thus we have
\begin{eqnarray}
\label{distbound}
D ( \sigma_i , \ket{ \psi_M } ) & \leq & D ( \sigma_i , \sigma^P_i ) +
D ( \sigma^P_i , \ket{ \psi_M } ) \\ \nonumber
& \leq &
( 1 - r_0^i - r_1^i )^{\frac{1}{2}} +
( \frac{\delta^2}{4} + ( \Delta s_0^i )^2 )^{\frac{1}{2}}\\\nonumber
& \leq &
\max_{j}\{( 1 - r_0^j - r_1^j )^{\frac{1}{2}} +
( \frac{\delta^2}{4} + ( \Delta s_0^j )^2 )^{\frac{1}{2}} \} \, .
\end{eqnarray}
This last expression is experimentally measurable and bounds the
separation between one of the evolved states and a maximally
superposed state, although the argument does not identify either state.

\section{Entanglement generation via candidate SWAP gates}

Consider now an experiment implementing a candidate SWAP gate
on a system of two exact or approximate qubits.  If we initialize
the system in the state $\ket{0}\ket{1}$, an exact exchange
interaction would evolve
the system to $\ket{1}\ket{0}$ via a maximally entangled state
of the form
\newline$\ket{\psi_M}
= \frac{1}{\sqrt{2}} (\ket{0}\ket{1} + e^{ i \phi} \ket{1}\ket{0})$.
We can consider the two qubits as defining a single effective qubit
in the two dimensional space with computational basis $\ket{0}\ket{1}$ and
$\ket{1}\ket{0}$.  If the system state is initially in this space,
it will, to good approximation, remain there provided that the
evolution is well approximated by an exchange interaction.
Our results for the NOT evolution of approximate qubits thus
apply directly.

Moreover, by carrying out appropriate
measurements, we can verify that an entangled state is
generated during the candidate SWAP evolution and give
figures of merit for the entanglement.
Define the projection $Q$ to be on the four dimensional
space defined by the two qubits.
A measurement of $Q$ on the state $\sigma_i$ would produce
the exact qubit state $\sigma_i^Q = Q \sigma_i Q / \Tr (Q \sigma_i )$
with probability $ p_Q = \sum_{l=0,1;m=0,1} p_{lm}^{i}$,
where $p_{lm}^i = \bra{l}\bra{m} \sigma_i \ket{l}\ket{m}$.
For any maximally entangled state $\ket{ \psi_M }$
we have
\begin{eqnarray} \max_i \bra{ \psi_M } \sigma_i^Q \ket{ \psi_M } & \geq &
\min_j \{ \frac{ p_{01}^j + p_{10}^j }{ p_{00}^j +
p_{10}^j + p_{01}^j + p_{11}^j } \} \times \nonumber \\
&& \max_k \{  \bra{ \psi_M } \sigma_k^P \ket{ \psi_M } \} \, .
\end{eqnarray}
As we can calculate both expressions on the right hand side, we
have a lower bound on the fidelity of a $\sigma_i^Q$
to a maximally entangled state.   If this is greater than one half,
the relevant $\sigma_i^Q$ is entangled, and as $\sigma_i^Q$ can
be obtained from $\sigma_i$ by local measurements, $\sigma_i$ is also
entangled.  The advantage of this procedure over
more efficient ways of measuring entanglement\cite{lewenstein,guhne,enttom,fiurasek} 
is that it requires no gates other than the one being tested.

\section{Discussion}

We have shown that, if a candidate NOT gate acts on an exact qubit,
the qubit must pass close to an equal weight superposition of
$|0\rangle$ and $|1\rangle$ if the NOT operation is demonstrated 
to work accurately on both basis states as inputs.  For an approximate 
qubit measurements are also needed half way through the NOT operation,
to check that the system has not wandered far out of the qubit subspace.
However, in both cases superposition can be inferred without
plotting oscillations in detail, given reasonable physical assumptions.

These physical assumptions should hold true in the types of 
experiment we consider. However, it should be stressed that they 
could in principle be violated given sufficient pathologies.  
To take an extreme example, one could imagine a single qubit 
experiment in which, unknown to the experimenter,
the environment happened to contain a second qubit, coupled to the
first by a fast but only infrequently acting swap gate.  The initially
prepared computational basis state in the first qubit could then be
exchanged with a maximally mixed state in the second, rotated via
a fortuitous NOT operation in the second qubit, and then exchanged
back into the first.  Applying our inequalities would lead to
the incorrect conclusion that the first qubit evolves through
a near-maximal superposition state. In fact, in this case,
such a state is indeed created, but in the second qubit lurking in
the environment. Of course, it is very unlikely indeed
that any experiment will accidentally incorporate a hidden
qubit or anything similar.

The arguments used to infer a superposition in a single qubit from
computational basis measurements can also
be used to infer entanglement in certain multi-qubit systems. For a
two qubit system arranged to effect a SWAP operation, demonstration
that this works well for the two inputs $\ket{0}\ket{1}$ and
$\ket{1}\ket{0}$ and measurements at the half way point (root SWAP)
showing that the system hasn't wandered far out of the subspace
spanned by these states are sufficient to infer that the system passed
close to a maximally entangled state. Fidelities can be bounded
and the existence of entanglement established, even for states well away
from being maximally entangled. Once again, all this works whether
the qubits are exact or approximate.

Clearly, for candidate condensed matter systems to be feasible qubits for
quantum computation, reliable single qubit gates will eventually need
to be constructed. This would remove the need to restrict to measurements 
in the computational basis, and would make tomography much simpler.  
However, state of the art is currently well short of this goal. 
At present, experimenters are trying to realise approximate one 
and two condensed matter qubit gates and test them.
In particular, recent superconducting experiments \cite{pashkin} show
that the coupling of two condensed matter qubits and investigations of
entanglement are now becoming possible. We believe that the results 
presented here should be extremely useful during the present 
investigative period. Our technique achieves less than full 
quantum state tomography would, but requires considerably 
less experimental effort, while providing a simple way of
inferring essential quantum behaviour in new candidate qubit 
experiments. Even in the longer term, our results should 
be relevant for fundamental investigations such as creating 
multi-qubit GHZ states or entangling two Schr\"{o}dinger cats, 
which may well be carried out on systems
where our measurement restrictions apply.

\vskip10pt
\leftline{\bf Acknowledgements} \qquad  This work was supported in part by the
European project EQUIP, IST-1999-11053. We thank Simon Benjamin,
Lucien Hardy, Brendon Lovett and Martin Plenio for valuable discussions.


\begin{thebibliography}{99}
\bibitem{example} The charge-phase qubit of ref. 7 can be biased so the 
measurable computational basis states are energy eigenstates (charge superpositions). 
The Coulomb interaction between two such qubits would provide a swapping
interaction between $\ket{0}\ket{1}$ and $\ket{1}\ket{0}$.
\end{thebibliography}
\end{document}